\documentclass[12pt]{article}
\usepackage{graphicx}
\def \b{{\cal B}}

\def \beq{\begin{equation}}
\def \beqn{\begin{eqnarray}}

\def \eeq{\end{equation}}
\def \eeqn{\end{eqnarray}}
\def \ca{{\cal A}}
\def \cn{Collaboration}
\def \ite{{\it et al.}}
\def \ob{\overline{B}^0}
\def \obs{\overline{B}^0_s}

\def \s{\sqrt{2}}

\def \st{\sqrt{3}}
\def \sx{\sqrt{6}}
\def \bu{{\bar u}}
\def \bd{{\bar d}}
\def \bs{{\bar s}}

\def \v#1#2{V_{#1#2}}

\begin{document}
\renewcommand{\thetable}{\Roman{table}}
\rightline{EFI-03-03-Rev}
\rightline{hep-ph/0302110}
\rightline{April 2003}
\bigskip
\bigskip
\centerline{\bf FINAL-STATE PHASES IN $B \to $ BARYON-ANTIBARYON DECAYS
\footnote{Submitted to Phys.~Rev.~D.}}
\bigskip
\centerline{\it Zumin Luo and Jonathan L. Rosner}
\centerline{\it Enrico Fermi Institute and Department of Physics}
\centerline{\it University of Chicago, Chicago, IL 60637}
\bigskip
\centerline{\bf ABSTRACT}
\medskip
\begin{quote}
The recent observation of the decay $\ob \to \Lambda_c^+ \bar p$
suggests that related decays may soon be visible at $e^+ e^-$
colliders.  It is shown how these decays can shed light on strong
final-state phases and amplitudes involving the spectator quark,
both of which are normally expected to be small in $B$ decays.
\end{quote}
\medskip
\leftline{\qquad PACS codes: 13.25.Hw, 11.30.Hv, 14.40.Nd, 13.75.Lb}
\bigskip

\centerline{\bf I. INTRODUCTION}
\bigskip

Phases in $B$ decays arising from final-state interactions are an
important gateway to the observation of direct CP violation.  The
pattern of decays to $D \pi$, $D^* \pi$, $D \rho$, and related states
has been elaborated recently by the CLEO \cite{PedlarD,Ahmed}, BaBar
\cite{BaDsK}, and Belle \cite{BeDro,BeDK,BeDsK} Collaborations.  Some
amplitudes for decays involving the weak subprocess $b \to c \bar u d$
obey isospin triangle relations.  In certain cases these triangles have
non-zero area, indicating non-zero final-state phases between different
contributing amplitudes \cite{CR}.  Some decays governed by the
Cabibbo-suppressed subprocess $b \to c \bar u s$ also involve amplitude
triangles with apparently non-zero area, though not yet at a statistically
significant level \cite{CR,Xing}.  One would expect this behavior if
flavor SU(3) is a good symmetry for $B$ decays.

The decays of $B$ mesons to charmed baryon -- charmless antibaryon pairs also
obey simple isospin relations and flavor-SU(3) regularities \cite{LW,SW}.
Models for these decays \cite{DTS,CZ,Jarfi,BD,CY} have been published which
allow estimates of their rates.  The recent observation of the decay $\ob \to
\Lambda_c^+ \bar p$ by the Belle Collaboration \cite{BeLp} with a branching
ratio $\b(\ob \to \Lambda_c^+ \bar p) = (2.19 ^{+0.56}_{-0.49} \pm 0.32 \pm
0.57) \times 10^{-5}$ indicates that such processes are within experimental
reach at existing $e^+ e^-$ colliders.
(Many early models \cite{DTS,CZ,Jarfi,BD} overestimated branching ratios to
baryon-antibaryon final states but contain useful theoretical techniques.)
The present paper indicates how these data may be useful in elaborating
final-state phases among different amplitudes contributing to the decays. 
It also indicates how one can test for suppression of decay amplitudes
involving the spectator quark.

We shall discuss the decomposition of $\ob \to \Lambda_c^+ \bar p$
and related decays into invariant amplitudes of flavor SU(3) in
Sec.\ II.  The triangles formed by these amplitudes, and their
significance for final-state interactions, are discussed in Sec.\
III.  We conclude with some experimental prospects in Sec.\ IV.
Conventions for the quark composition of baryons are given in the
Appendix.

\bigskip
\centerline{\bf II.  INVARIANT AMPLITUDES OF FLAVOR SU(3)}
\bigskip

The weak Hamiltonian giving rise to the subprocess $b \to c \bar u
d$ transforms as the $I = 1, I_3 = -1$ member of an octet of
flavor SU(3).  The $\overline B$ mesons $b \bar q$ ($\bar q=-\bar
u,\bar d,\bar s$) form a $3^*$. (Recall that $(-\bar u, \bar d)$
is an isodoublet.) Thus the SU(3) representations of the initial
state are those in the product
\beq \label{eqn:To}
3^* \times 8 = 3^* + 6 + 15^*~~~.
\eeq
The $\Lambda_c^+ = c[ud]$ belongs to a flavor-SU(3) antitriplet
($3^*$) along with the $\Xi_c^+ = c[su]$ and the $\Xi_c^0 =
c[sd]$. The brackets indicate antisymmetry with respect to flavor.
For decays to a final state of a $3^*$ charmed baryon and an octet
antibaryon, all three representations in Eq.\ (\ref{eqn:To})
occur.  Hence there must be three independent invariant amplitudes
of flavor SU(3) characterizing such decays. Similarly, in the
Cabibbo-suppressed decays governed by $b \to c \bar u s$, the weak
Hamiltonian transforms as the strange charged isodoublet member of
a flavor octet, so the invariant amplitudes are the same.

Charmed baryons belonging to a flavor-SU(3) sextet ($6$) also have
been seen, consisting of an isotriplet $\Sigma_c^{++} =
cuu,~\Sigma_c^+ = c(ud), \Sigma_c^0 = cdd$, an isodoublet
${\Xi'}_c^+ = c(us),~{\Xi'}_c^0 = c(ds)$, and an isosinglet
$\Omega_c^0 = css$.  The parentheses indicate symmetry with
respect to flavor. Similarly, one can consider not only octet but
also (anti)decuplet antibaryons. In Table \ref{tab:reps} we
summarize the SU(3) representations that contribute to each
class of decays.

\begin{table}
\caption{Invariant amplitudes in the direct channel contributing
to $\overline B \to$ (charmed baryon) + (antibaryon) decays via
the subprocess $b \to c \bar u d$ or $b \to c \bar u s$.
\label{tab:reps}}
\begin{center}
\begin{tabular}{l c c} \hline \hline
\quad Charmed baryon & $3^*$ & 6 \\
Antibaryon           &       &   \\ \hline
8                    & $3^* + 6 + 15^*$ & $3^* + 6 + 15^*$ \\
$10^*$               &        6         & $3^* + 15^*$ \\ \hline \hline
\end{tabular}
\end{center}
\end{table}

An economical tensor notation was utilized by Savage and Wise to
describe these processes \cite{SW}.  We illustrate with the $3^* +
8$ final state.  We use subscripts to denote the components of a
$3^*$ representation of $SU(3)_f$ and use superscripts to denote
the components of a $3$ representation. The $\overline B$ mesons,
in a $3^*$ representation as mentioned, can then be written as
$(-B^-, \ob, \obs) \equiv B_i$. The charmed baryons in a $3^*$
representation can be expressed as $(-\Xi_c^0, \Xi_c^+,
\Lambda_c^+) \equiv (\Xi_c)_i$. The octet of charmless baryons, on
the other hand, can be represented by a two-index tensor:
\beq
 N^i_j \equiv \left( \begin{array}{ccc}
-\Sigma^0/\sqrt{2}+\Lambda/\sqrt{6} & \Sigma^+ & p
\\ -\Sigma^- & \Sigma^0/\sqrt{2}+\Lambda/\sqrt{6} & n \\ -\Xi^- & \Xi^0 &
 -2\Lambda/\sqrt{6} \end{array} \right)~~~.
\eeq
The weak Hamiltonian responsible for the Cabibbo-favored quark
subprocess $b \to c d {\bar u}$ and the Cabibbo-suppressed $b \to
c s {\bar u}$, belonging to an $SU(3)_f$ octet as mentioned above, can
similarly be written as
\beq
H^i_j \sim (d{\bar u})V_{ud} + (s{\bar u})V_{us} = \left(
\begin{array}{ccc} 0 & 0 & 0
\\ V_{ud} & 0 & 0 \\ V_{us} & 0 & 0 \end{array} \right)~~~,
\eeq
where $V_{ud}$ and $V_{us}$ are the Cabibbo-Kobayashi-Maskawa
(CKM) matrix elements: \\ $V_{us}/V_{ud} \simeq  \lambda \simeq 0.2256$.
The effective Hamiltonian for the
decays $B \to \Xi_c {\overline N}$ can be written in terms of
invariant amplitudes $\alpha, \beta$ and $\gamma$ \cite{SW}:
\beq \label{eqn:SW} H_{\rm eff} = \alpha \Xi_c^i N^j_i H^k_j B_k
+\beta \Xi_c^i H^j_i N^k_j B_k +\gamma \Xi_c^i B_i H^j_k N^k_j~~~,
\eeq
where we sum over repeated indices. Expanding the sum would give
us the amplitudes for the relevant processes. (Remember to
multiply each amplitude by $(-1)^{n_{\bar u}}$, where $n_{\bar u}$
is the number of $\bar u$ quarks in the antibaryon.)

Two equivalent notations are helpful to visualize possible
relations among invariant amplitudes.  The second is particularly
relevant when certain dynamical assumptions are made.

(1) The process $3^* \times 8 \to 3^* \times 8$ in the crossed channel reads
\beq
3^* \times 3 \to 1 + 8_D + 8_F \to 8 \times 8~~~,
\eeq
where $D$ and $F$ denote the two ways of coupling an octet to a pair of octets.
The singlet $S$ and octet amplitudes $D$ and $F$ (suitably normalized) are
related to $\alpha$, $\beta$, and $\gamma$ by
\beq
\alpha = D + F~~,~~~\beta = D - F~~,~~~\gamma = S - \frac{2}{3} D~~~.
\eeq
The $S,~D,~F$ notation is that (aside from normalization) used by
Li and Wu \cite{LW}.  In Table \ref{tab:xreps} we summarize the SU(3)
representations that contribute to each class of decays, including also
sextet charmed baryons and antidecuplet antibaryons.  We see, of course,
that the number of invariant amplitudes is the same as in the direct
channel.

\begin{table}
\caption{Invariant amplitudes in the crossed channel contributing
to $\overline B \to$ (charmed baryon) + (antibaryon) decays via
the subprocess $b \to c \bar u d$ or $b \to c \bar u s$.
\label{tab:xreps}}
\begin{center}
\begin{tabular}{l c c} \hline \hline
\quad Charmed baryon & $3^*$ & 6 \\
Antibaryon           &       &   \\ \hline
8                    & $1 + 8_D + 8_F$ & $8_D + 8_F + 10$ \\
$10^*$               &        8        & $8 + 10$ \\ \hline \hline
\end{tabular}
\end{center}
\end{table}

\begin{figure}
\begin{center}
\includegraphics[height=2.2in]{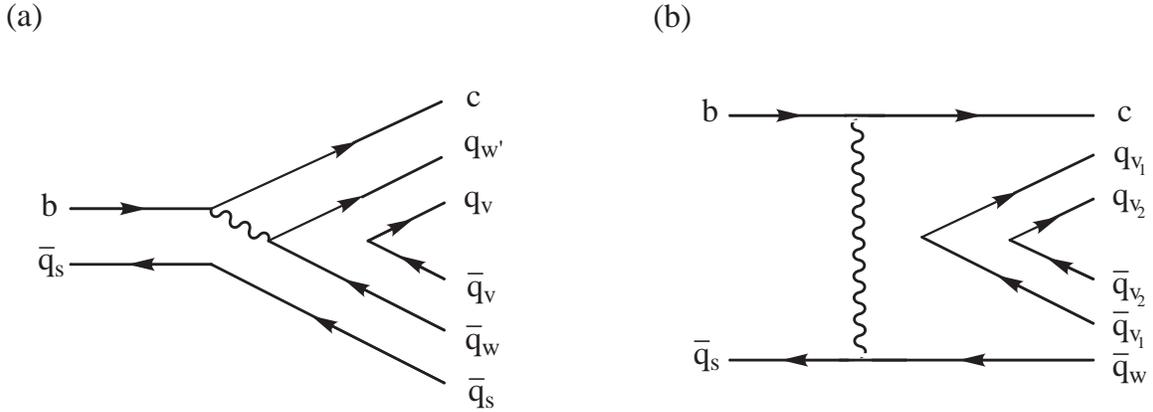}
\caption{Diagrams for $\overline B$ $\to$ a charmed  baryon and an
antibaryon. (a) Color-suppressed diagram. $q_{w'} = d$ for
Cabibbo-favored decays and $q_{w'} = s$ for Cabibbo-suppressed
decays; $\bar q_w=\bar u$. (b) Exchange diagram. $\bar q_s = \bar
d$ for  Cabibbo-favored decays and $\bar q_s = \bar s$ for
Cabibbo-suppressed decays; $\bar q_w=\bar u$. \label{fig:decay}}
\end{center}
\end{figure}

(2) A topological expansion of amplitudes \cite{Chau,GHLR} yields
three invariant amplitudes of which two are associated with the
subprocess $b \to c d \bar u$ or $b \to c s \bar u$, with an
additional light quark-antiquark pair produced from the vacuum
[Fig.~\ref{fig:decay}(a)], and one is associated with the exchange
process $b \bar d \to c \bar u$ or $b \bar s \to c \bar u$, in
which two such pairs are produced from the vacuum
[Fig.~\ref{fig:decay}(b)].  We call the first two amplitudes $a_1$
and $a_2$ and the third amplitude $a_E$ (to denote exchange).
Explicit definitions of these amplitudes are given below. Consider
the amplitudes for $\overline B$ to decay to 6 quarks
($c~q_{w'}~q_v$ and $\bar q_s$~$\bar q_w$~$\bar q_v$) via
color-suppressed processes as shown in Fig.~\ref{fig:decay}(a).
With the $c$ quark staying at the top, there are 2 permutations
for \{$q_{w'}~q_v$\} and 6 permutations for \{$\bar q_s$~$\bar
q_w$~$\bar q_v$\}. Thus there are 12 possible color-suppressed
diagrams contributing to a specific amplitude. The amplitudes of
the 12 diagrams are denoted by $A_{i j k}^{l m}$, where $lm$ is a
permutation of \{$w' ~v$\} and $ijk$ is a permutation of
\{$s~w~v$\}. The color-suppressed amplitude for $\overline B$ to
decay to a charmed baryon and an antibaryon is then a weighted sum
of the 12 amplitudes, with the weights being the products of the
coefficient of $cq_lq_m$ in the quark composition of the charmed
baryon and that of $\bar q_i\bar q_i\bar q_k$ in the quark
composition of the antibaryon. It turns out that each
color-suppressed amplitude is a linear combination of $a_1$ and
$a_2$, with
\beqn a_1 & = & \frac{1}{2}(A_{[sw]v}^{[w'v]} + A_{[sv]w}^{[w'v]}) \\
a_2 & = & \frac{1}{2}(A_{[ws]v}^{[w'v]} + A_{[wv]s}^{[w'v]})~~~. \eeqn
Here $A_{[ij]k}^{[lm]} \equiv (A_{ijk}^{lm}-A_{jik}^{lm}) -
(A_{ijk}^{ml}-A_{jik}^{ml})$ and $1/2$ is merely a normalization
factor. Similarly, $E_{ijk}^{lm}$ is used to denote the amplitude
for ${\overline B}$ to decay to 6 quarks via an exchange process
as shown in Fig.~\ref{fig:decay}(b). Here $lm$ is a permutation of
\{$v_1 ~v_2$\} and $ijk$ is a permutation of \{$w~v_1~v_2$\}.
Since the two quark-antiquark pairs ($q_{v_1}\bar q_{v_1}$ and
$q_{v_2}\bar q_{v_2}$) are both produced from the vacuum, $a_E$
should not depend on the ordering of $v_1$ and $v_2$. One finds
that all exchange amplitudes for $\overline B$ to decay to a
charmed baryon and an antibaryon are multiples of
\beq a_E = \frac{1}{2}(E_{[v_1v_2]w}^{[v_1v_2]} +
E_{[wv_2]v_1}^{[v_1v_2]} - E_{[wv_1]v_2}^{[v_1v_2]})~~~. \eeq
The topological decompositions of the amplitudes are presented in
Table~\ref{tab:To}. They are in agreement with those obtained from
Eq.~(\ref{eqn:SW}) if we set
\beq a_1 =-\gamma,~~~a_2=-\beta,~~~a_E = \alpha+\gamma~~~. \eeq
In particular, if processes involving the spectator quark are
suppressed, as has been argued for heavy-quark decays (see, e.g.,
the discussion in \cite{GHLR}), one expects $|a_E| \ll
|a_1|,~|a_2|$, and hence an approximate symmetry \beq \alpha = -
\gamma~~~. \eeq We shall explore the consequences of this relation
in the next Section.
%
\begin{table}
\caption{$SU(3)_f$ predictions of the amplitudes for $\overline B
\to$ an SU(3) $3^*$ charmed baryon and an octet antibaryon. CF =
Cabibbo-favored, CS = Cabibbo-suppressed. \label{tab:To}}
\begin{center}
\begin{tabular}{l c l c}
\hline \hline
CF Decay & Amplitude & CS Decay & Amplitude \\
\hline ${\overline B}^0 \to \Lambda_c^+ {\bar p}$ & $a_1+a_E$ &
${\overline B}_s^0 \to \Xi_c^+ {\overline \Sigma}^-$ & $\lambda (a_1+a_E)$ \\
${\overline B}_s^0 \to \Xi_c^0 {\overline \Xi}^0$ & $-a_2$ &
${\overline B}^0 \to \Xi_c^0 {\bar n}$ & $-\lambda a_2$ \\
${\overline B}_s^0 \to \Lambda_c^+ {\overline \Sigma}^-$ &
$-a_1$ & ${\overline B}^0 \to \Xi_c^+ {\bar p}$ & $-\lambda a_1$ \\
${\overline B}^0 \to \Xi_c^0 {\overline \Sigma}^0$ &
$-(a_1+a_2+a_E)/\sqrt{2}$ & ${\overline
B}_s^0 \to \Xi_c^0 {\overline \Sigma}^0$ & $-\lambda (a_1+a_E)/\sqrt{2}$ \\
${\overline B}^0 \to \Xi_c^0 {\overline \Lambda}$ &
$(a_1-a_2+a_E)/\sqrt{6}$ & ${\overline
B}_s^0 \to \Xi_c^0 {\overline \Lambda}$ & $\lambda (a_1+2a_2+a_E)/\sqrt{6}$ \\
${\overline B}^0 \to \Xi_c^+ {\overline \Sigma}^-$ & $a_E$ &
${\overline
B}_s^0 \to \Lambda_c^+ {\bar p}$ & $\lambda a_E$ \\
$B^- \to \Xi_c^0 {\overline \Sigma}^-$ & $-(a_1+a_2)$ & $B^-
\to \Xi_c^0 {\bar p}$ & $-\lambda (a_1+a_2)$ \\
\hline \hline
\end{tabular}
\end{center}
\end{table}
%

More generally, the topological amplitudes contributing to each
type of process are summarized in Table \ref{tab:gphs}. For decays
to $6 + 8$, both $q_{w'}q_v$ and $q_{v_1}q_{v_2}$ are symmetrized
and therefore
\beqn b_1 & = & \frac{1}{2}(A_{[sw]v}^{(w'v)} + A_{[sv]w}^{(w'v)}) \\
b_2 & = & \frac{1}{2}(A_{[ws]v}^{(w'v)} + A_{[wv]s}^{(w'v)}) \\
b_E & = & \frac{1}{2}(E_{[wv_1]v_2}^{(v_1v_2)} +
E_{[wv_2]v_1}^{(v_1v_2)} )~~~ \label{eqn:bE}, \eeqn
where $A_{[ij]k}^{(lm)} \equiv (A_{ijk}^{lm}-A_{jik}^{lm}) +
(A_{ijk}^{ml}-A_{jik}^{ml})$ and $E_{[ij]k}^{(lm)}$ is defined in
a similar way. Note that if $q_{v_1}$ and $q_{v_2}$ are identical,
only one term in Eq.~(\ref{eqn:bE}) contributes. For decays to
$3^* + 10^*$, there is no exchange diagram since $q_{v_1}$ and
$q_{v_2}$ are antisymmetrized in a $3^*$ charmed baryon but $\bar
q_{v_1}$ and $\bar q_{v_2}$ are symmetrized in a $10^*$
antibaryon; and
\beq c = A_{(swv)}^{[w'v]}/\sqrt{2} \equiv \sum_{\sigma}
(A_{\sigma\{swv\}}^{w'v} - A_{\sigma\{swv\}}^{vw'})/\sqrt{2}~~~,
\eeq
where the sum runs over all permutations $\sigma$ of \{$s~w~v$\}.
For decays to $6 + 10^*$,
\beq d = A_{(swv)}^{(w'v)} \equiv \sum_{\sigma}
(A_{\sigma\{swv\}}^{w'v} + A_{\sigma\{swv\}}^{vw'})~~~, \eeq
and $d_E = E_{(wv_1v_2)}^{(v_1v_2)}$ is defined in a similar
fashion. In Tables \ref{tab:So}--\ref{tab:Sd} we summarize the
corresponding amplitudes for decays to $6 + 8$, $3^* + 10^*$, and
$6 + 10^*$, respectively.  These are equivalent to the
decompositions presented in Ref.\ \cite{SW}, but we find the
present notation convenient for seeing what happens when we assume
that the exchange amplitudes are small. We do not show the
amplitudes for Cabibbo-suppressed decays (which can be looked up
in \cite{SW}), since these decays generally involve $\Xi'_c$,
$\Omega_c$ or ${\overline B}_s^0$, none of which is easy to
observe or produce in experiments. Furthermore, the branching
ratios for these decays are expected to be only a few percent of
those for the Cabibbo-favored ones.

\begin{table}
\caption{Invariant amplitudes in a topological expansion for $\overline B
\to$ (charmed baryon) + (antibaryon) decays via the subprocess $b \to c
\bar u d$ or $b \to c \bar u s$. \label{tab:gphs}}
\begin{center}
\begin{tabular}{l c c} \hline \hline
\quad Charmed baryon & $3^*$ & 6 \\
Antibaryon           &       &   \\ \hline
8    & $a_1,~a_2,~a_E$ & $b_1,~b_2,~b_E$ \\
$10^*$ &   $c$     & $d,~d_E$ \\
\hline \hline
\end{tabular}
\end{center}
\end{table}

\begin{table}
\caption{$SU(3)_f$ predictions of the amplitudes for $\overline B
\to (6$ charmed baryon $+$ octet antibaryon). Only Cabibbo-favored
decays are shown. \label{tab:So}}
\begin{center}
\begin{tabular}{l c l c} \hline \hline
Decay & Amplitude & Decay & Amplitude \\ \hline $\ob \to
\Sigma_c^0 \bar n$ & $-(b_2 + b_E)$ $^{a)}$ &
  $B^- \to \Sigma_c^0 \bar p$ & $-(b_1+b_2)$ $^{b)}$\\
$\ob \to \Sigma_c^+ \bar p$ & $(b_E-b_1)/\s$ &
  $B^- \to {\Xi'}_c^0 \overline \Sigma^-$ & $(b_1+b_2)/\s$ \\
$\ob \to {\Xi'}_c^0 \overline \Lambda$ & $(b_2-b_1+3 b_E)/(2 \st)$
&
  $\overline{B}_s^0 \to {\Xi'}_c^0 \overline \Xi^0$ & $b_2/\s$ \\
$\ob \to {\Xi'}_c^0 \overline \Sigma^0$ & $(b_1+b_2+b_E)/2$ &
  $\overline{B}_s^0 \to \Sigma_c^+ \overline \Sigma^-$ & $b_1/\s$ \\
$\ob \to {\Xi'}_c^+ \overline \Sigma^-$ & $-b_E/\s$ &
  $\overline{B}_s^0 \to \Sigma_c^0 \overline \Sigma^0$ & $-b_1/\s$ \\
$\ob \to \Omega_c^0 \overline \Xi^0$ & $b_E$ &
  $\overline{B}_s^0 \to \Sigma_c^0 \overline \Lambda$ & $(b_1+2b_2)/\sx$ \\
\hline \hline
\end{tabular}
\end{center}
\footnotesize{a). The branching ratio for this mode is predicted
to be ${\cal O}(10^{-7} \sim 10^{-6})$ in a pole model \cite{CY}.\\
b). A branching ratio of $(0.45^{+0.26}_{-0.19} \pm 0.07 \pm 0.12)
\times 10^{-4}$ is measured for $B^- \to \Sigma_c^0 \bar p$ by the
Belle Collaboration \cite{Gabyshev:2002zq}. This sets a 90\% CL
upper limit $0.93 \times 10^{-4}$, to be compared with the 90\% CL
upper limit $0.8 \times 10^{-4}$ set by the CLEO Collaboration
\cite{Dytman:2002yd}. A prediction based on the pole model of Ref.\
\cite{CY} agrees with these limits.}
\end{table}

\begin{table}
\caption{$SU(3)_f$ predictions of the amplitudes for $\overline B
\to (3^*$ charmed baryon $+$ antidecuplet antibaryon)
(Cabibbo-favored decays). Note that $\ca(\ob \to \Xi_c^+ \overline
\Sigma^{*-}) = 0$. \label{tab:Td}}
\begin{center}
\begin{tabular}{l c l c} \hline \hline
Decay & Amplitude & Decay & Amplitude \\ \hline $\ob \to
\Lambda_c^+ \overline \Delta^-$ & $- c/\st$ &
  $\ob \to \Xi_c^0 \overline \Sigma^{*0}$ & $c/\sx$ \\
$B^- \to \Lambda_c^+ \overline \Delta^{--}$ & $- c$ $^{a)}$ &
  $B^- \to \Xi_c^0 \overline \Sigma^{*-}$ & $c/\st$ \\
$\overline{B}_s^0 \to \Lambda_c^+ \overline \Sigma^{*-}$ &
$-c/\st$ & $\overline{B}_s^0 \to \Xi_c^0 \overline \Xi^{*0}$ & $c/\st$ \\
\hline \hline
\end{tabular}
\end{center}
\footnotesize{ a). Branching ratios of $(1.87^{+0.43}_{-0.40} \pm
0.28 \pm 0.49) \times 10^{-4}$ and $(2.4 \pm 0.6^{+0.19}_{-0.17}
\pm 0.6) \times 10^{-4}$ are observed for $B^- \to \Lambda_c^+
\bar p \pi^-$ by the Belle \cite{Gabyshev:2002zq} and CLEO \cite{Dytman:2002yd}
Collaborations, respectively.  Since
$\overline \Delta^{--}$ decays almost exclusively to $\bar
p \pi^-$, the branching ratio for $B^- \to \Lambda_c^+ \overline
\Delta^{--}$ should be less than ${\cal B}(B^- \to \Lambda_c^+
\bar p \pi^-)$. The pole model or Ref.\ \cite{CY} predicts ${\cal B}(B^- \to
\Lambda_c^+ \overline \Delta^{--}) = 1.9 \times 10^{-5}$.}
\end{table}

\begin{table}
\caption{$SU(3)_f$ predictions of the amplitudes for $\overline B
\to (6$ charmed baryon $+$ antidecuplet antibaryon)
(Cabibbo-favored decays). \label{tab:Sd}}
\begin{center}
\begin{tabular}{l c l c} \hline \hline
Decay & Amplitude & Decay & Amplitude \\ \hline $\ob \to
\Sigma_c^{++} \overline \Delta^{--}$ & $-d_E$ $^{a)}$ &
   $B^- \to \Sigma_c^+ \overline \Delta^{--}$ & $d/\s$ \\
$\ob \to \Sigma_c^+ \overline \Delta^{-}$ & $(2d_E+d)/\sx$ &
  $B^- \to \Sigma_c^0 \overline \Delta^-$ & $-d/\st$ \\
$\ob \to \Sigma_c^0 \overline \Delta^0$ & $-(d_E + d)/\st$ &
  $B^- \to {\Xi'}_c^0 \overline \Sigma^{*-}$ & $-d/\sx$ \\
$\ob \to {\Xi'}_c^+ \overline \Sigma^{*-}$ & $(2/3)^{1/2}d_E$ &
  $\overline{B}_s^0 \to \Sigma_c^+ \overline \Sigma^{*-}$ & $d/\sx$ \\
$\ob \to {\Xi'}_c^0 \overline \Sigma^{*0}$ & $-(2d_E+d)/(2 \st)$ &
  $\overline{B}_s^0 \to \Sigma_c^0 \overline \Sigma^{*0}$ & $-d/\sx$ \\
$\ob \to \Omega_c^0 \overline \Xi^{*0}$ & $-d_E/\st$ &
  $\overline{B}_s^0 \to {\Xi'}_c^0 \overline \Xi^{*0}$ & $-d/\sx$ \\
\hline \hline
\end{tabular}
\end{center}
\footnotesize{a). The branching ratio for $\ob \to \Sigma_c^{++}
\overline \Delta^{--}$ should be less than that for $\ob \to \Sigma_c^{++}
\bar p \pi^-$. The latter is measured to be $(2.38^{+0.63}_{-0.55} \pm
0.41 \pm 0.62) \times 10^{-4}$ and $(3.7 \pm 0.8 \pm 0.7 \pm 1.0) \times
10^{-4}$ by the Belle \cite{Gabyshev:2002zq} and CLEO
\cite{Dytman:2002yd} Collaborations, respectively.}
\end{table}

\bigskip
\centerline{\bf III. TRIANGLE RELATIONS}
\bigskip

In all the processes we consider, the charmed baryon has spin 1/2.  Since
the decaying particle has spin 0, and parity is not conserved in the
decay, there are two independent amplitudes, labeled by the helicity of
the charmed baryon. The following triangle relations are valid for
each. The parity-conserving (PC) and parity-violating (PV) amplitudes are
linear combinations of the two helicity amplitudes. In some models (see,
e.g., \cite{Jarfi}), one of the amplitudes (e.g., PV) may be absent or
suppressed with respect to the other. In the absence of final-state
phases, one can show that the triangle formed by the square roots of three
decay rates has zero area if and only if the PC and PV amplitudes for the
three decay processes form similar triangles. Indeed, zero final-state
phases and similar PC and PV triangles are two necessary and sufficient
conditions for the triangle formed by the square roots of three decay
rates to have zero area. The proof is given below. 

Suppose that $s_i^2 = |c_i|^2 + |v_i|^2$ ($i=1, 2, 3$) are the decay rates
for three processes, with $c_i$ and $v_i$ being the PC and PV amplitudes,
respectively. Assuming that these amplitudes satisfy triangle relations
$c_1 + c_2 = c_3$ and $v_1 + v_2 = v_3$, we have
\begin{eqnarray*}
s_3^2 & = & s_1^2 + s_2^2 + 2 {\rm Re} (c_1 c_2^* + v_1 v_2^*) \\
& \le & s_1^2 + s_2^2 + 2 (|c_1||c_2| + |v_1||v_2|) \\
& \le & s_1^2 + s_2^2 + 2 \sqrt{|c_1|^2 + |v_1|^2}\sqrt{|c_2|^2+|v_2|^2}
\\ & = & (s_1 + s_2)^2~~~,
\end{eqnarray*}
where the second inequality is due to the Cauchy-Schwarz
inequality. Obviously, the equality $s_3 = s_1 + s_2$ holds if and only if
there are no relative phases both between $c_1$ and $c_2$ and between
$v_1$ and $v_2$, and the relation $|c_1|/|c_2| = |v_1|/|v_2|$ is
satisfied.  One then has $|c_1|/|v_1| = |c_2|/|v_2| = |c_3|/|v_3|$.

In what follows we shall assume that, by studying decay distributions, one has
been able to separate out the individual rates for parity-conserving and
parity-violating transitions, or the individual rates for charmed baryon
helicities $\pm 1/2$.  In the case of amplitude equalities (rather than
triangle relations), total rates as well as individual ones will of course be
equal.
\bigskip

\leftline{\bf A.  $3^* + 8$ final states.}
\bigskip

\begin{figure}
\begin{center}
\includegraphics[height=3.5in]{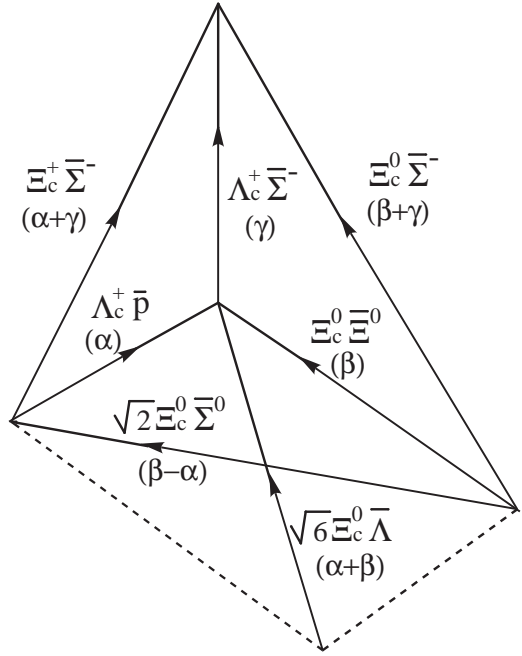}
\caption{Triangles for $\ca(\overline{B}^0 \to \Lambda_c^+ \bar p) =
\alpha$ and related amplitudes described in Table~\ref{tab:To}.
Note that $\alpha = a_1 + a_E$, $\beta = - a_2$ and $\gamma = -
a_1$. \label{fig:amps}}
\end{center}
\end{figure}

The Cabibbo-favored amplitudes of Table~\ref{tab:To} are denoted by arrows in
Fig.~\ref{fig:amps}.  For each helicity or partial wave, three independent
complex amplitudes will be specified completely, up to an irrelevant
overall phase, by five lengths of these vectors, leaving two
predictions for rates.  There will be a discrete ambiguity
corresponding to the folding of two adjacent triangles about their
common side.  (We do not show the corresponding figure for
Cabibbo-suppressed decays.)  We now discuss some individual
triangle relations associated with this construction.  These
triangles, if shown to have non-zero area, will indicate non-zero
relative final-state phases between their contributing amplitudes.

As a consequence of the isospin of the weak Hamiltonian for $b \to
c \bar u d$, two invariant isospin amplitudes, with $I = 1/2$ and
$I = 3/2$, govern $\overline{B} \to \Xi_c \overline \Sigma$.  The
three decay processes then obey a triangle relation:
\beq \ca(B^- \to \Xi^0_c \overline \Sigma^-) = \s \ca(\ob \to
\Xi^0_c \overline \Sigma^0) + \ca(\ob \to \Xi^+_c \overline
\Sigma^-)~~~. \eeq
This relation is somewhat challenging in view of the need to
reconstruct the $\overline \Sigma^0$ through its $\overline \Lambda
\gamma$ decay.  However, it involves only non-strange $B$ mesons, which
are the focus of current studies at $e^+ e^-$ colliders.

Three triangle relations involve the observed $\ob \to \Lambda_c^+
\bar p$ decay:
\beq \s \ca(\ob \to \Lambda^+_c \bar p) + \ca(\ob \to \Xi^0_c
\overline \Sigma^0) = \st \ca(\ob \to \Xi^0_c \overline
\Lambda)~~~, \eeq \beq \ca(\ob \to \Lambda_c^+ \bar p) + \ca(\obs
\to \Xi_c^0 \overline \Xi^0) = \sx \ca(\ob \to \Xi_c^0 \overline
\Lambda)~~~, \eeq \beq \ca(\ob \to \Lambda_c^+ \bar p) + \ca(\obs
\to \Lambda_c^+ \overline \Sigma^-) = \ca(\ob \to \Xi_c^+
\overline \Sigma^-)~~~. \eeq
The first one is particularly useful since it involves only $\ob$
decays. The last two relations involve the detection of a $\obs$
decay, requiring either a dedicated run at KEKB or PEP-II
(currently running below $B^0_s \obs$ threshold) or an experiment at
a hadron collider.

In the Cabibbo-suppressed sector two isospin relations stem from
the $I = 1/2$, $I_3 = -1/2$ nature of the weak Hamiltonian:
\beq \ca(B^- \to \Xi_c^0 \bar p) = \ca(\ob \to \Xi_c^0 \bar n) +
\ca(\ob \to \Xi_c^+ \bar p)~~~, \eeq \beq \ca(\obs \to \Xi_c^+
\overline \Sigma^-) = - \s \ca(\obs \to \Xi_c^0 \overline
\Sigma^0)~~~. \eeq
The first of these involves only non-strange $B$'s and no
$\overline \Sigma^0$'s. Two additional triangle relations may be
written, both involving $\obs$ decays.  Since these involve
Cabibbo-suppressed decays of the less easily produced $\obs$, the
corresponding triangles may not be so easy to construct.

\bigskip
\leftline{\bf B.  $6+8$ final states.}
\bigskip

The isospin triangles in these processes are
\beq \ca(B^- \to \Sigma_c^0 \bar p) = \s \ca(\ob \to \Sigma_c^+
\bar p) + \ca(\ob \to \Sigma_c^0 \bar n)~~~, \eeq
which involves an antineutron, and
\beq \ca(B^- \to {\Xi'}_c^0 \overline \Sigma^-) = \s \ca(\ob \to
{\Xi'}_c^0 \overline \Sigma^0) + \ca(\ob \to {\Xi'}_c^+ \overline
\Sigma^-)~~~, \eeq
which involves the ${\Xi'}_c$ states. These were not observed
until quite recently \cite{Jessop:1998wt} since they decay to
$\Xi_c \gamma$. A simple isospin relation \beq \ca(\obs \to
\Sigma_c^+ \overline \Sigma^-) = - \ca(\obs \to \Sigma_c^0
\overline \Sigma^0) \eeq involves $\obs$ decays.  Several
amplitude triangles not involving isospin can be formed from the
relations for $6 + 8$ decays, but they involve particles which are
not especially easy to produce ($\obs$) or detect (${\Xi'}_c$).

There are several ways to check whether the exchange amplitude
$b_E$ is much smaller than $b_1$ or $b_2$.  For example, the decay
$\ob \to {\Xi'}_c^+ \overline \Sigma^-$ occurs only via the
exchange amplitude, so it would be suppressed in comparison with
the other decays to ${\Xi'}_c \overline \Sigma$.  Similarly, the
decay $\ob \to \Omega_c^0 \overline \Xi^0$ would be suppressed.
If, indeed, $b_E$ is found to be suppressed, a useful amplitude
triangle based on the two independent amplitudes $b_1$ and $b_2$
could be formed:
\beq 2 \st \ca(\ob \to {\Xi'}_c^0 \overline \Lambda) + \ca(B^- \to
\Sigma_c^0 \bar p) = 2 \s \ca(\ob \to \Sigma_c^+ \bar p)~~~. \eeq
Other such triangles can also be formed, but they generally
involve $\obs$ decays.

\bigskip
\leftline{\bf C.  $3^* + 10^*$ final states.}
\bigskip

Here a single amplitude describes all decays.  The relation
\beq \ca(B^- \to \Lambda_c^+ \overline \Delta^{--}) = \st \ca(B^0
\to \Lambda_c^+ \overline \Delta^-) \eeq
is a consequence of the pure isospin $(I=3/2)$ of the final state.
The decays $\overline{B} \to \Xi_c \overline \Sigma^*$ involve
both $I = 1/2$ and $I = 3/2$, but these amplitudes are related to
one another since $\ob \to \Xi_c^+ \overline \Sigma^{*-}$ is
forbidden. This process could only have proceeded via an exchange
amplitude, but the final charmed baryon is antisymmetric in its
light quarks, which cannot couple to the symmetrized quarks in the
final antidecuplet antibaryon.  Thus the isospin relation
\beq \ca(B^- \to \Xi_c^0 \overline \Sigma^{*-}) = \s \ca(\ob \to
\Xi_c^0 \overline \Sigma^{*0}) + \ca(\ob \to \Xi_c^+ \overline
\Sigma^{*-})~~~ \eeq
is implemented as
\beq \ca(B^- \to \Xi_c^0 \overline \Sigma^{*-}) = \s \ca(\ob \to
\Xi_c^0 \overline \Sigma^{*0})~~~. \eeq
There are no triangle relations, and no tests for a vanishing
exchange amplitude since it never contributes in the first place.

\bigskip
\leftline{\bf D.  $6 + 10^*$ final states.}
\bigskip

There are a number of isospin triangles involving the charge
states of $\overline{B} \to \Sigma_c \overline \Delta$.  One
example for which detection of final states may be particularly
favorable is
\beq \ca(\ob \to \Sigma_c^{++} \overline \Delta^{--}) + \st
\ca(B^- \to \Sigma_c^0 \overline \Delta^-) = \st \ca(\ob \to
\Sigma_c^0 \overline \Delta^0)~~~. \eeq
Another useful relation involves the two charge states of $B^- \to
\Sigma_c \overline \Delta$:
\beq  \s \ca(\ob \to \Sigma_c^+ \overline \Delta^{--}) + \st
\ca(B^- \to \Sigma_c^0 \overline \Delta^-) = 0~~~. \eeq
In order that the isospin triangles have non-zero area, both $d$
and $d_E$ must be nonvanishing and have a nontrivial relative
phase. A good test for $d_E = 0$ is to check whether the decay
$\ob \to \Sigma_c^{++} \overline \Delta^{--}$ is suppressed in
comparison with other $\overline{B} \to \Sigma_c \overline \Delta$ decays.
When $d_E = 0$, all the rates for processes in Table \ref{tab:Sd} are
either zero or related to one another by simple factors.

Another isospin triangle involving the charge states of
$\ob \to \Xi_c \overline \Sigma^*$ is
\beq \ca(B^- \to {\Xi'}_c^0 \overline \Sigma^{*-}) = \s \ca(\ob
\to {\Xi'}_c^0 \overline \Sigma^{*0}) + \ca(\ob \to {\Xi'}_c^+
\overline \Sigma^{*-})~~~. \eeq
However, experimentally it is not easy to construct.

\bigskip
\centerline{\bf IV.  DISCUSSION AND SUMMARY}
\bigskip

The recent observation of a two-body baryon-antibaryon $B$ decay
\cite{BeLp} is likely to be the first in a series of such decays.
We have shown that these processes are capable of providing
information on two main questions which have been of interest in
$B$ meson decays for some years:  (1) Are there significant
final-state interaction phases between different decay amplitudes
characterized by the same weak phases?  (2) Are processes
involving the spectator quark (such as the exchange amplitudes
described here by the suffix $E$) suppressed in comparison with
other amplitudes in which the spectator does not enter into the
weak Hamiltonian?  We have described a number of tests of both
these questions which may be feasible in the near future.
In particular, if amplitude triangles formed of total rates for three processes
appear to have zero area, we have shown that relative final-state phases must
vanish {\it and} that parity-conserving and parity-violating transition
amplitudes must be in the same proportion in all three processes.
Tests for the smallness of exchange amplitudes can be performed by several
comparisons of rates in Tables \ref{tab:To}, \ref{tab:So}, and \ref{tab:Sd}.
Other tests may require separation of helicity amplitudes before being
fully implemented.

Given the value of the observed branching ratio for $\ob \to \Lambda_c^+
\bar p$ \cite{BeLp}, which was based on an integrated luminosity
of 78.2 fb$^{-1}$, several times the present data sample may be
needed to see some of the related decay modes, but the triangle
construction in Fig.~\ref{fig:amps} suggests that at least some
other decay modes to a charmed baryon and an octet antibaryon may
be observable with comparable branching ratios. 
Combined with the predictions of Ref.\ \cite{CY} and the assumption of
suppression of the exchange amplitudes, Table \ref{tab:So}
indicates that a few other processes (such as $\ob \to \Sigma_c^+
\bar p$, $B^- \to {\Xi'}_c^0 \overline \Sigma^-$,
$\overline{B}_s^0 \to \Sigma_c^+ \overline \Sigma^-$ and
$\overline{B}_s^0 \to \Sigma_c^0 \overline \Sigma^0$) may have
branching ratios of about the same order as the already observed decay
$B^- \to \Sigma_c^0 \bar p$. Another decay in Table \ref{tab:Td},
$B^- \to \Lambda_c^+ \overline \Delta^{--}$, should also be
observable if its branching ratio is of order $10^{-5}$ as
predicted by the pole model of Ref.\ \cite{CY}.

\bigskip
\centerline{\bf ACKNOWLEDGEMENTS}
\bigskip

This work was supported in part by the United States Department of Energy,
High Energy Physics Division, under Contract No.\ DE-FG02-90ER-40560.

\bigskip
\centerline{\bf APPENDIX: QUARK COMPOSITION OF BARYONS}
\bigskip

In our convention ($\Xi_c^0$, $\Xi_c^+$), ($\bar p$, $\bar n$),
($\overline \Sigma^-$, $\overline \Sigma^0$, $\overline
\Sigma^+$), ($\overline \Xi^0$, $\overline \Xi^+$), ($\overline
\Delta^{--}$, $\overline \Delta^-$, $\overline \Delta^0$,
$\overline \Delta^+$), ($\overline \Sigma^{*-}$, $\overline
\Sigma^{*0}$, $\overline \Sigma^{*+}$) and ($\overline \Xi^{*0}$,
$\overline \Xi^{*+}$) are in iso-multiplets. We recall that $I_- u
= d$, $I_- \bar d = - \bar u$. Our convention for the $\overline
B$ mesons is: $B^- = - b \bar u$, $\overline B^0 = b \bar d$,
$\overline B_s^0 = b \bar s$.
\bigskip

\noindent I) Antitriplet charmed baryons:
\begin{eqnarray*}
\Lambda_c^+ & = & (c u d - c d u)/\s \\
\Xi_c^+ & = & (c s u - c u s)/\s \\
\Xi_c^0 & = & (c s d - c d s)/\s
\end{eqnarray*}
II) Sextet charmed baryons:
\begin{eqnarray*}
\Sigma_c^{++} & = & c u u \\
\Sigma_c^+ & = & (c u d + c d u)/\s \\
\Sigma_c^0 & = & c d d \\
{\Xi'}_c^+ & = & (c u s + c s u)/\s \\
{\Xi'}_c^0 & = & (c d s + c s d)/\s \\
\Omega_c^0 & = & c s s
\end{eqnarray*}
III) Octet antibaryons:
\begin{eqnarray*}
\bar p & = & (\bu \bd \bu - \bd \bu \bu)/\s \\
\bar n & = & (\bd \bu \bd - \bu \bd \bd)/\s \\
\overline \Sigma^- & = & (\bs \bu \bu - \bu \bs \bu)/\s \\
\overline \Sigma^0 & = & (\bu \bs \bd - \bs \bu \bd + \bd \bs \bu - \bs \bd \bu)/2 \\
\overline \Sigma^+ & = & (\bs \bd \bd - \bd \bs \bd)/\s \\
\overline \Xi^0 & = & (\bu \bs \bs - \bs \bu \bs)/\s \\
\overline \Xi^+ & = & (\bs \bd \bs - \bd \bs \bs)/\s \\
\overline \Lambda & = & (2 \bu \bd \bs - 2 \bd \bu \bs - \bd \bs
\bu + \bs \bd \bu - \bs \bu \bd + \bu \bs \bd)/\sqrt{12}
\end{eqnarray*}
IV) Antidecuplet antibaryons:
\begin{eqnarray*}
\overline \Delta^{--}  & = & - \bu \bu \bu \\
\overline \Delta^- & = & (\bu \bu \bd + \bu \bd \bu + \bd \bu \bu)/\st \\
\overline \Delta^0 & = & -(\bu \bd \bd + \bd \bu \bd + \bd \bd \bu)/\st \\
\overline \Delta^+  & = & \bd \bd \bd \\
\overline \Sigma^{*-} & = & (\bu \bu \bs + \bu \bs \bu + \bs \bu \bu)/\st \\
\overline \Sigma^{*0} & = & -(\bu \bd \bs + \bu \bs \bd + \bd \bu \bs + \bd \bs \bu + \bs \bu \bd + \bs \bd \bu)/\sqrt{6} \\
\overline \Sigma^{*+} & = & (\bd \bd \bs + \bd \bs \bd + \bs \bd \bd)/\st \\
\overline \Xi^{*0} & = & -(\bu \bs \bs + \bs \bu \bs + \bs \bs \bu)/\st \\
\overline \Xi^{*+} & = & (\bd \bs \bs + \bs \bd \bs + \bs \bs \bd)/\st \\
\overline \Omega^+ & = & \bs \bs \bs
\end{eqnarray*}
%

\def \ajp#1#2#3{Am.~J.~Phys.~{\bf#1}, #2 (#3)}
\def \apny#1#2#3{Ann.~Phys.~(N.Y.) {\bf#1}, #2 (#3)}
\def \app#1#2#3{Acta Phys.~Polonica {\bf#1}, #2 (#3)}
\def \arnps#1#2#3{Ann.~Rev.~Nucl.~Part.~Sci.~{\bf#1}, #2 (#3)}
\def \art{and references therein}
\def \b97{{\it Beauty '97}, Proceedings of the Fifth International
Workshop on $B$-Physics at Hadron Machines, Los Angeles, October 13--17,
1997, edited by P. Schlein}
\def \carg{{\it Masses of Fundamental Particles -- Carg\`ese 1996}, edited by
M. L\'evy \ite, NATO ASI Series B:  Physics Vol.~363 (Plenum, New York, 1997)}
\def \cmp#1#2#3{Commun.~Math.~Phys.~{\bf#1}, #2 (#3)}
\def \cmts#1#2#3{Comments on Nucl.~Part.~Phys.~{\bf#1}, #2 (#3)}
\def \corn93{{\it Lepton and Photon Interactions:  XVI International
Symposium, Ithaca, NY August 1993}, AIP Conference Proceedings No.~302,
ed.~by P. Drell and D. Rubin (AIP, New York, 1994)}
\def \cp89{{\it CP Violation,} edited by C. Jarlskog (World Scientific,
Singapore, 1989)}
\def \dpff{{\it The Fermilab Meeting -- DPF 92} (7th Meeting of the
American Physical Society Division of Particles and Fields), 10--14
November 1992, ed. by C. H. Albright \ite~(World Scientific, Singapore,
1993)}
\def \dpf94{DPF 94 Meeting, Albuquerque, NM, Aug.~2--6, 1994}
\def \efi{Enrico Fermi Institute Report No. EFI}
\def \el#1#2#3{Europhys.~Lett.~{\bf#1}, #2 (#3)}
\def \epjc#1#2#3{Eur.~Phys.~J.~C {\bf#1}, #2 (#3)}
\def \f79{{\it Proceedings of the 1979 International Symposium on Lepton
and Photon Interactions at High Energies,} Fermilab, August 23-29, 1979,
ed.~by T. B. W. Kirk and H. D. I. Abarbanel (Fermi National Accelerator
Laboratory, Batavia, IL, 1979}
\def \hb87{{\it Proceeding of the 1987 International Symposium on Lepton
and Photon Interactions at High Energies,} Hamburg, 1987, ed.~by W. Bartel
and R. R\"uckl (Nucl. Phys. B, Proc. Suppl., vol. 3) (North-Holland,
Amsterdam, 1988)}
\def \ib{{\it ibid.}~}
\def \ibj#1#2#3{{\it ibid.}~{\bf#1}, #2 (#3)}
\def \ichep72{{\it Proceedings of the XVI International Conference on High
Energy Physics}, Chicago and Batavia, Illinois, Sept. 6--13, 1972,
edited by J. D. Jackson, A. Roberts, and R. Donaldson (Fermilab, Batavia,
IL, 1972)}
\def \ijmpa#1#2#3{Int.~J.~Mod.~Phys.~A {\bf#1}, #2 (#3)}
\def \ite{{\it et al.}}
\def \jmp#1#2#3{J.~Math.~Phys.~{\bf#1}, #2 (#3)}
\def \jpg#1#2#3{J.~Phys.~G {\bf#1}, #2 (#3)}
\def \lkl87{{\it Selected Topics in Electroweak Interactions} (Proceedings
of the Second Lake Louise Institute on New Frontiers in Particle Physics,
15--21 February, 1987), edited by J. M. Cameron \ite~(World Scientific,
Singapore, 1987)}
\def \KEK#1{{\it Flavor Physics} (Proceedings of the Fourth International
Conference on Flavor Physics, KEK, Tsukuba, Japan, 29--31 October 1996),
edited by Y. Kuno and M. M. Nojiri, Nucl.~Phys.~B Proc.~Suppl.~{\bf 59},
#1 (1997)}
\def \ky85{{\it Proceedings of the International Symposium on Lepton and
Photon Interactions at High Energy,} Kyoto, Aug.~19-24, 1985, edited by M.
Konuma and K. Takahashi (Kyoto Univ., Kyoto, 1985)}
\def \mpla#1#2#3{Mod.~Phys.~Lett.~A {\bf#1}, #2 (#3)}
\def \nc#1#2#3{Nuovo Cim.~{\bf#1}, #2 (#3)}
\def \nima#1#2#3{Nucl.~Instr.~Meth.~A {\bf#1}, #2 (#3)}
\def \np#1#2#3{Nucl.~Phys.~{\bf#1}, #2 (#3)}
\def \npbps#1#2#3{Nucl.~Phys.~B (Proc.~Suppl.) {\bf#1}, #2 (#3)}
\def \pisma#1#2#3#4{Pis'ma Zh.~Eksp.~Teor.~Fiz.~{\bf#1}, #2 (#3) [JETP
Lett. {\bf#1}, #4 (#3)]}
\def \pl#1#2#3{Phys.~Lett.~{\bf#1}, #2 (#3)}
\def \plb#1#2#3{Phys.~Lett.~B {\bf#1}, #2 (#3)}
\def \pr#1#2#3{Phys.~Rev.~{\bf#1}, #2 (#3)}
\def \pra#1#2#3{Phys.~Rev.~A {\bf#1}, #2 (#3)}
\def \prd#1#2#3{Phys.~Rev.~D {\bf#1}, #2 (#3)}
\def \prl#1#2#3{Phys.~Rev.~Lett.~{\bf#1}, #2 (#3)}
\def \prp#1#2#3{Phys.~Rep.~{\bf#1}, #2 (#3)}
\def \ptp#1#2#3{Prog.~Theor.~Phys.~{\bf#1}, #2 (#3)}
\def \rmp#1#2#3{Rev.~Mod.~Phys.~{\bf#1}, #2 (#3)}
\def \rp#1{~~~~~\ldots\ldots{\rm rp~}{#1}~~~~~}
\def \si90{25th International Conference on High Energy Physics, Singapore,
Aug. 2-8, 1990}
\def \slc87{{\it Proceedings of the Salt Lake City Meeting} (Division of
Particles and Fields, American Physical Society, Salt Lake City, Utah,
1987), ed.~by C. DeTar and J. S. Ball (World Scientific, Singapore, 1987)}
\def \slac89{{\it Proceedings of the XIVth International Symposium on
Lepton and Photon Interactions,} Stanford, California, 1989, edited by M.
Riordan (World Scientific, Singapore, 1990)}
\def \smass82{{\it Proceedings of the 1982 DPF Summer Study on Elementary
Particle Physics and Future Facilities}, Snowmass, Colorado, edited by R.
Donaldson, R. Gustafson, and F. Paige (World Scientific, Singapore, 1982)}
\def \smass90{{\it Research Directions for the Decade} (Proceedings of the
1990 Summer Study on High Energy Physics, June 25 -- July 13, Snowmass,
Colorado), edited by E. L. Berger (World Scientific, Singapore, 1992)}
\def \stone{{\it B Decays}, edited by S. Stone (World Scientific,
Singapore, 1994)}
\def \tasi90{{\it Testing the Standard Model} (Proceedings of the 1990
Theoretical Advanced Study Institute in Elementary Particle Physics,
Boulder, Colorado, 3--27 June, 1990), edited by M. Cveti\v{c} and P.
Langacker (World Scientific, Singapore, 1991)}
\def \vanc{29th International Conference on High Energy Physics, Vancouver,
23--31 July 1998}
\def \yaf#1#2#3#4{Yad.~Fiz.~{\bf#1}, #2 (#3) [Sov.~J.~Nucl.~Phys.~{\bf #1},
#4 (#3)]}
\def \zhetf#1#2#3#4#5#6{Zh.~Eksp.~Teor.~Fiz.~{\bf #1}, #2 (#3) [Sov.~Phys.
-- JETP {\bf #4}, #5 (#6)]}
\def \zpc#1#2#3{Zeit.~Phys.~C {\bf#1}, #2 (#3)}

\end{document}